\documentstyle[11pt,epsfig]{article}
\title{\bf Modified dispersion relations in extra dimensions}
\author{A. S. Sefiedgar$^1$\thanks{e-mail: a-sefiedgar@sbu.ac.ir},\quad
K. Nozari$^2$\thanks{email: knozari@umz.ac.ir}\,\,\, and \,\, H. R.
Sepangi$^1$\thanks{email: hr-sepangi@sbu.ac.ir}
\\ {\small $^1$Department of Physics, Shahid Beheshti University G. C., Evin,
Tehran 19839, Iran}
\\ {\small $^2$Department of Physics, Faculty of Basic Sciences,
University of Mazandaran}\\{\small P.O. Box 47416-95447, Babolsar,
Iran}}
\begin{document}
\maketitle
\begin{abstract}
It has recently been shown that the thermodynamics of a FRW universe
can be fully derived using the generalized uncertainty principle
(GUP) in extra dimensions as a primary input. There is a
phenomenologically close relation between the GUP and Modified
Dispersion Relations (MDR). However, the form of the MDR in theories
with extra dimensions is as yet not known. The purpose of this
letter is to derive the MDR in extra dimensional scenarios. To
achieve this goal, we focus our attention on the thermodynamics of a
FRW universe within a proposed MDR in an extra dimensional model
universe. We then compare our results with the well-known results
for the thermodynamics of a FRW universe in an extra dimensional GUP
setup. The result shows that the entropy functionals calculated in
these two approaches are the same, pointing to a possible conclusion
that these approaches are equivalent. In this way, we derive the MDR
form in a model universe with extra dimensions that would have
interesting implications on the construction of the ultimate quantum
gravity scenario.
\end{abstract}
\vspace{2cm}
\section{Introduction}
A common feature of all promising candidates for quantum gravity is
the existence of a minimal observable length \cite{1,2,3,4,5}.
Modified dispersion relations (MDR) and the generalized uncertainty
principle (GUP) are two approaches to phenomenologically incorporate
this finite resolution of the space-time points within the
theoretical framework of the standard model. In fact, MDR and GUP
are common features to all candidates of quantum gravity models. In
particular, in the study of loop quantum gravity and models based on
non-commutative geometry, there has been strong interest in some
modifications of the energy-momentum dispersion relation
\cite{6,7,8,9,10}. On the other hand, the generalized uncertainty
principle has been considered in string theory and in the models
based on noncommutative geometry \cite{1,2,3,4,5}. The MDR and GUP
essentially affect the thermodynamics of physical systems at energy
scales within the realm of quantum gravity. Thus the exact form of
the GUP and MDR could essentially lead one to a deeper understanding
of the ultimate quantum gravity proposal.

Our goal here is to deduce the form of the MDR in a model universe
with extra dimensions through the application of the machinery of
black hole thermodynamics to the universe as a unique physical
system. In this respect, we look at the status of this connection in
two main frameworks.

Firstly, it has been shown recently that one can generalize the
well-known approach to black hole thermodynamics to study
thermodynamics of the Universe as a unique physical system through
its \emph{apparent horizon}. In this way, it has been revealed that
there is a deep connection between thermodynamics and gravity
through the laws of black hole thermodynamics \cite{11,111,12}. This
connection has also been realized for a FRW universe \cite{13}. In a
FRW universe, which is the subject of the present study, one would
replace the event horizon of a black hole by the apparent horizon of
a FRW spacetime. One then assumes that the apparent horizon has an
associated entropy $S$ defined as $S=\frac{A}{4G}$ and a temperature
$T$ defined as $T=\frac{1}{2 \pi \tilde{r}_{_A}}$, where $G$ is the
gravitational constant, $A$ is the area of the apparent horizon and
$\tilde{r}_{_A}$ is the radius of the apparent horizon. In this
viewpoint, the first law of thermodynamics, that is $dE=TdS$, can be
translated into the language of the Friedmann equation. The first
law of thermodynamics plays a crucial role in different theories
such as Einstein gravity, Gauss-Bonnet gravity, Lovelock gravity and
various braneworld scenarios \cite{14,15,16}. Hence one can infer a
deep connection between gravity and thermodynamics in this viewpoint
\cite{17}.

Secondly, the Hawking radiation at the vicinity of a black hole
event horizon \cite{111,18} or at the apparent horizon of a FRW
spacetime \cite{19} is another important issue worth discussing. The
Hawking radiation is a quantum mechanical effect in the classical
background of a black hole spacetime or FRW geometry. Therefore,
quantum mechanics, gravitational theory and thermodynamics meet each
other when the physics of black holes and FRW spacetime  are
concerned. A thorough study of black holes with a size comparable to
the Planck scale or a FRW universe in the Planck era needs quantum
gravity considerations. In other words, a complete quantum gravity
scenario is required \cite{17} to handle these important issues.
Since GUP and MDRs are common features to all quantum gravity
approaches, one may apply them to obtain thermodynamics of a FRW
universe or black hole in the small length scale, or equivalently in
the high energy regime. In the past few years, the GUP and MDR have
been applied to modify the Hawking radiation and  Bekenstein-Hawking
entropy of black holes \cite{20,21,22,23,24,25,26,28}.
Thermodynamical properties of a FRW universe as another important
example of a quantum gravity regime has been studied within the GUP
formalism in arbitrary dimensions \cite{17}. The thermodynamics of
the FRW universe has also been considered within the MDR formalism
in a 4-dimensional spacetime \cite{27}.

The study of thermodynamics when extra dimensions are present is an
interesting aspect of the discussions above, on which we shall
concentrate in this work. One may consider the role played by extra
dimensions in the study of thermodynamics of the FRW universe within
the MDR framework and compare the results  with that when the GUP is
used.  We therefore start by deriving the FRW universe
thermodynamics using the well-known form of the generalized
uncertainty principle in extra dimensions. Then, motivated by the
form of the GUP in an extra dimensional scenario, we suggest a
modified dispersion relation. We then use this form of the MDR to
find the FRW universe entropy. Since MDR and GUP are different
manifestations of the same concept (existence of a minimal length
scale or a maximum momentum) in the quantum gravity theory, we
expect that the entropy calculated in these two frameworks should be
the same. In fact, we expect the results of application of these two
approaches to the issue of apparent horizon thermodynamics should be
the same at least in their functional form (and not necessarily in
their numerical coefficients). The assumption behind this
expectation is that GUP and MDR are phenomenologically two (though
seemingly different) faces of an underlying quantum gravity
proposal. By comparing the results deduced from these two
approaches, we fix the functional form of the MDR suggested for a
higher dimensional spacetime. In other words, consistency of
thermodynamics in these two approaches constrains the form of the
MDR suggested for a higher dimensional spacetime. Since we know the
exact form of the extra dimensional GUP from various works (see
\cite{17,28} for instance), we deduce the form of MDR in a spacetime
with extra dimensions. Such a study has been lacking in the
literature and would be useful in the construction of a successful
theory of quantum gravity.

\section{Entropy of a FRW universe}
We consider a $(n+1)$-dimensional FRW universe with the following
line element
\begin{eqnarray}\label{1}
ds^2=-dt^2+a^2(t)\left(\frac{dr^2}{1-kr^2}+r^2d\Omega_{n-1}^2\right),
\end{eqnarray}
where $d\Omega_{n-1}^2$ denotes the line element of a
$(n-1)$-dimensional unit sphere, $a(t)$ is the scale factor of our
universe and $k$ is the spatial curvature constant. Using the
notation $\tilde{r}=ar$, the radius of the apparent horizon can be
written as
\begin{eqnarray}\label{2}
\tilde{r}_{_A}=\frac{1}{\sqrt{H^2+k/{a^2}}},
\end{eqnarray}
where $H\equiv\frac{\dot{a}}{a}=\frac{da/dt}{a}$ is the Hubble
parameter. We assume that the apparent horizon has an associated
entropy $S$ and temperature $T$ defined as
\begin{eqnarray}\label{3}
S=\frac{A}{4G}\,, \qquad T=\frac{1}{2 \pi \tilde{r}_{_A}},
\end{eqnarray}
respectively. Here, $A$ is the apparent horizon area and $G$ is the
gravitational constant. It has been shown that the first law of
thermodynamics
\begin{eqnarray}\label{4}
dE=TdS,
\end{eqnarray}
reproduces the Friedmann equation in this framework \cite{17}. With
these preliminaries, we are now ready to obtain the entropy of a FRW
universe in the frameworks of the GUP and MDR models in a  universe
with extra dimensions. We then compare our results obtained in these
two approaches to constrain the form of the proposed MDR in extra dimensional scenarios.

\subsection{The GUP }

We consider the following GUP
\begin{eqnarray}\label{5}
\delta x \delta p\geq 1+\alpha^2 l_p^2 \delta p^2,
\end{eqnarray}
where $l_p$ is the Planck length depending on the
dimensionality and fundamental energy scale of the extra dimensional
model universe we are interested in and $\alpha$ is a
dimensionless real constant \cite{17,28}. It is easy to show that
\begin{eqnarray}\label{6}
\delta p\geq \frac{1}{\delta x} \left [\frac{\delta
x^2}{2\alpha^2l_p^2} -\frac{\delta
x^2}{2\alpha^2l_p^2}\sqrt{1-\frac{4\alpha^2l_p^2}{\delta
x^2}}\right] =\frac{1}{\delta x}\Psi\left(\delta x^2\right),
\end{eqnarray}
where
\begin{eqnarray}\label{7}
\Psi(\delta x^2)=\frac{\delta x^2}{2\alpha^2l_p^2} -\frac{\delta
x^2}{2\alpha^2l_p^2}\sqrt{1-\frac{4\alpha^2l_p^2}{\delta x^2}},
\end{eqnarray}
shows the departure of the GUP from the standard uncertainty
principle. Here we assume that a particle with energy $dE$ is
absorbed or radiated via the apparent horizon. The energy of this
particle may be identified by $dE\simeq\delta p$ ( with $c=1$)
\cite{21}. Within the Heisenberg uncertainty principle,
 $\delta p\geq 1/{\delta x}$, one may find from equations (\ref{3}) and (\ref{4}) that
\begin{eqnarray}\label{8}
dA=\frac{4G}{T}dE\simeq\frac{4G}{T}\frac{1}{\delta x}.
\end{eqnarray}
Incorporating the effect of the GUP via the inclusion of $\Psi$, one
finds
\begin{eqnarray}\label{9}
dA_{\Psi}=\frac{4G}{T}dE\simeq\frac{4G}{T}\frac{1}{\delta
x}\Psi\left(\delta x^2\right).
\end{eqnarray}
Using equation (\ref{8}), we find
\begin{eqnarray}\label{10}
dA_{\Psi}\simeq\Psi(\delta x^2)d A.
\end{eqnarray}
The position uncertainty $\delta x$ of the particle crossing through
the apparent horizon can be chosen as its Compton wavelength which
is proportional to the inverse of the Hawking temperature. Hence one
can write \cite{22,28}
\begin{eqnarray}\label{11}
\delta
x\simeq2\tilde{r}_A=2\left(\frac{A}{n\Omega_n}\right)^{\frac{1}{n-1}}\,,
\end{eqnarray}
where $\Omega_n$ is the volume of an n-dimensional unit sphere. Now
one obtains $\Psi(\delta x^2)$ as a function of the area of the
apparent horizon as follows
\begin{eqnarray}\label{12}
\Psi(A)=\frac{2}{\alpha^2
l_p^2}\left(\frac{A}{n\Omega_n}\right)^{\frac{2}{n-1}}
\left(1-\sqrt{1-\alpha^2 l_p^2
\left(\frac{n\Omega_n}{A}\right)^{\frac{2}{n-1}}}\right).
\end{eqnarray}
At $\alpha=0$, one may use the Taylor expansion to obtain
\begin{eqnarray}\label{13}
\Psi(A)=1+\frac{\alpha^2l_p^2}{4}\left(\frac{n\Omega_n}{A}\right)^{\frac{2}{n-1}}
+\frac{\alpha^4l_p^4}{8}\left(\frac{n\Omega_n}{A}\right)^{\frac{4}{n-1}}
+\frac{15
\alpha^6l_p^6}{192}\left(\frac{n\Omega_n}{A}\right)^{\frac{6}{n-1}}+\cdots
.
\end{eqnarray}
Considering only the terms up to the sixth power of the Planck
length (without loss of generality in conclusion), one finds
\begin{eqnarray}\label{14}
dA_\Psi \simeq\left[1+\frac{\alpha^2l_p^2}{4}\left(\frac{n\Omega_n}{A}
\right)^{\frac{2}{n-1}}+\frac{\alpha^4l_p^4}{8}\left(\frac{n\Omega_n}{A}\right)^{\frac{4}{n-1}}
+\frac{15
\alpha^6l_p^6}{192}\left(\frac{n\Omega_n}{A}\right)^{\frac{6}{n-1}}\right]d
A.
\end{eqnarray}
Integrating equation (\ref{14}) gives the modified area of the
apparent horizon, $A_\Psi$. One may then substitute $A_\Psi$ in
$S_\Psi=\frac{A_\Psi}{4G}$ to find the GUP-corrected entropy. For
$n=3$, the GUP-corrected entropy of the FRW universe will be
\begin{eqnarray}\label{15}
S_\Psi \simeq\frac{A}{4G}+\frac{1}{4}\alpha^2
l_p^2\left(\frac{3\Omega_3}{4G}\right)\ln{\frac{A}{4G}}
-\frac{1}{8}\alpha^4
l_p^4\left(\frac{3\Omega_3}{4G}\right)^2\frac{4G}{A}
-\frac{15}{384}\alpha^6
l_p^6\left(\frac{3\Omega_3}{4G}\right)^3\left(\frac{4G}{A}\right)^2.
\end{eqnarray}
For $n=4$ the corrected entropy is
\begin{eqnarray}\label{16}
S_\Psi \simeq\frac{A}{4G}+\frac{3}{4}\alpha^2
l_p^2\left(\frac{4\Omega_4}{4G}\right)^{\frac{2}{3}}
\left(\frac{A}{4G}\right)^{\frac{1}{3}}-\frac{3}{8}\alpha^4 l_p^4
\left(\frac{4\Omega_4}{4G}\right)^{\frac{4}{3}}\left(\frac{4G}{A}\right)^{\frac{1}{3}}
-\frac{15}{192}\alpha^6
l_p^6\left(\frac{4\Omega_4}{4G}\right)^2\left(\frac{4G}{A}\right).
\end{eqnarray}
Similarly, for $n=5$ the GUP-corrected entropy is
\begin{eqnarray}\label{17}
S_\Psi \simeq\frac{A}{4G}+\frac{1}{2}\alpha^2
l_p^2\left(\frac{5\Omega_5}{4G}
\right)^{\frac{1}{2}}\left(\frac{A}{4G}\right)^{\frac{1}{2}}+\frac{1}{8}\alpha^4
l_p^4
\left(\frac{5\Omega_5}{4G}\right)\ln{\frac{A}{4G}}-\frac{30}{192}\alpha^6
l_p^6
\left(\frac{5\Omega_5}{4G}\right)^{\frac{3}{2}}\left(\frac{4G}{A}\right)^{\frac{1}{2}}.
\end{eqnarray}
As can be seen from the above relations, the logarithmic correction
term only appears in a FRW spacetime with even number of dimensions.
In other words, only for odd $n$, the GUP-corrected entropy contains
a logarithmic correction term. The impact of the generalized
uncertainty principle on the black hole entropy was previously
considered  in \cite{28} where the emergence of the logarithmic
correction term was restricted to the even-dimensional spacetimes.
If the mysterious existence of the logarithmic correction term in
the black hole or FRW universe entropy  is proven rigorously in
future, it will impose stringent constraints on the number of the
spacetime dimensions.

\subsection{The MDR }

As was indicated previously, the functional form of a typical MDR in
a model universe with extra dimensions has not been studied yet.
Ordinarily, an understanding of the concept of MDR in a
$(n+1)$-dimensional spacetime may require  one to expect that its
fundamental length scale, $L_p$, should be different from that of
the 4-dimensional case in order to account for the existence of the
extra dimensions. What we propose to do in this regard is to
postulate that the modified dispersion relation in a model universe
with extra dimensions can be written in the same way as in
$4$-dimensions \cite{30}
\begin{eqnarray}\label{18}
(\overrightarrow{p})^2=f(E,m;L_p)\simeq E^2-\mu^2+\alpha
L_p^2E^4+\alpha' L_p^4E^6+\alpha'' L_p^6E^8+{\cal
O}\left(L_P^8E^{10}\right),
\end{eqnarray}
where $L_p$ is the Planck length which depends on the dimensions of
the spacetime we are interested in and $f$ is the function that
gives the exact dispersion relation. On the right hand side we have
assumed the applicability of a Taylor-series expansion for $E\ll
\frac{1}{L_p}$. The coefficients $\alpha_i$ may take different
values in different quantum gravity proposals. Note that $m$ is the
rest energy of the particle and the mass parameter $\mu$ on the
right hand side is directly related to the rest energy. However,
$\mu\neq m$ if $\alpha_i$'s do not all vanish. As we have emphasized
previously, to incorporate quantum gravitational effects,
thermodynamics of the FRW universe  should be modified. Of course
MDR may provide a perturbation framework for this modification.
Using equation (\ref{18}) and applying a Taylor expansion, we find
\begin{eqnarray}\label{19}
dp \simeq dE\left[1+\frac{3}{2}\alpha
L_p^2E^2+\left(\frac{5}{2}\alpha'-\frac{5}{8}\alpha^2
\right)L_p^4E^4+\left(\frac{7}{2}\alpha''-\frac{7}{4}\alpha \alpha'+
\frac{21}{48}\alpha^3\right)L_p^6E^6\right]\,,
\end{eqnarray}
where we have kept only the terms up to the sixth power of the
Planck length, without loss of generality in conclusion. Some
manipulations will then lead to
\begin{eqnarray}\label{20}
dE \simeq dp\left[1-\frac{3}{2}\alpha
L_p^2E^2+\left(-\frac{5}{2}\alpha'+\frac{23}{8}\alpha^2
\right)L_p^4E^4+\left(-\frac{7}{2}\alpha''+\frac{37}{4}\alpha
\alpha'- \frac{273}{48}\alpha^3\right)L_p^6E^6\right].
\end{eqnarray}
To first order, assuming $E\sim \delta E$, we may apply the standard
uncertainty formulae, $\delta E\geq \frac{1}{\delta x}$\,  and \,
$\delta p\geq\frac{1}{\delta x}$\, to obtain
\begin{eqnarray}\label{21}
dE\geq \frac{1}{\delta x}\left[1-\frac{3}{2}\alpha
L_p^2\frac{1}{\delta x^2}+
\left(-\frac{5}{2}\alpha'+\frac{23}{8}\alpha^2\right)L_p^4\frac{1}{\delta
x^4}+ \left(-\frac{7}{2}\alpha''+\frac{37}{4}\alpha
\alpha'-\frac{273}{48}\alpha^3 \right)L_p^6\frac{1}{\delta
x^6}\right].
\end{eqnarray}
This relation can be rewritten as
\begin{eqnarray}\label{22}
dE\geq \frac{1}{\delta x}\Phi\left(\delta x^2\right),
\end{eqnarray}
where by definition
\begin{eqnarray}\label{23}
\Phi(\delta x^2)=1-\frac{3}{2}\alpha L_p^2\frac{1}{\delta x^2}+
\left(-\frac{5}{2}\alpha'+\frac{23}{8}\alpha^2\right)L_p^4\frac{1}{\delta
x^4}+ \left(-\frac{7}{2}\alpha''+\frac{37}{4}\alpha
\alpha'-\frac{273}{48}\alpha^3 \right)L_p^6\frac{1}{\delta x^6}\,\,.
\end{eqnarray}
The corresponding relation in the standard framework is given by
equation (\ref{8}). We note that in these equations the
trace of extra dimensions is encoded in $L_{p}$ which depends on the
dimensionality of spacetime manifold. Taking into
account the effect of MDR, we find
\begin{eqnarray}\label{24}
dA_{\Phi}=\frac{4G}{T}dE\simeq\frac{4G}{T}\frac{1}{\delta
x}\Phi\left(\delta x^2\right).
\end{eqnarray}
This means that
\begin{eqnarray}\label{25}
dA_{\Phi} \simeq \Phi(\delta x^2)d A.
\end{eqnarray}
Now, using equation (\ref{11}), we can write
\begin{eqnarray}\label{26}
dA_{\Phi} \simeq \Phi(A)d A,
\end{eqnarray}
where
\begin{eqnarray}\label{27}
\Phi(A)&=&1-\frac{3}{8}\alpha L_p^2
\left(\frac{n\Omega_n}{A}\right)^{\frac{2}{n-1}}+
\left(-\frac{5}{32}\alpha'+\frac{23}{128}\alpha^2\right)L_p^4
\left(\frac{n\Omega_n}{A}\right)^{\frac{4}{n-1}}+\nonumber\\
&+&\left(-\frac{7}{128}\alpha''+\frac{37}{256}\alpha \alpha'-
\frac{273}{3072}\alpha^3\right)L_p^6\left(\frac{n\Omega_n}{A}\right)^{\frac{6}{n-1}}.
\end{eqnarray}
Integrating equation (\ref{26}) and substituting the result in
equation $S_{\Phi}=\frac{A_\Phi}{4G}$, we derive the entropy of the
FRW universe for a model universe with extra dimensions. For $n=3,
4, 5$ the entropy of the FRW universe become
\begin{eqnarray}\label{28}
S_\Phi&\simeq&\frac{A}{4G}-\frac{3}{8}\alpha
L_p^2\left(\frac{3\Omega_3}{4G}\right)
\ln{\frac{A}{4G}}-\left(-\frac{5}{32}\alpha'+\frac{23}{128}\alpha^2\right)
L_p^4\left(\frac{3\Omega_3}{4G}\right)^2\frac{4G}{A}
\nonumber\\&-&\frac{1}{2}\left(-\frac{7}{128}\alpha''+\frac{37}{256}\alpha
\alpha'-\frac{273}{3072}\alpha^3\right)L_p^6\left(\frac{3\Omega_3}{4G}
\right)^3\left(\frac{4G}{A}\right)^2\,,
\end{eqnarray}

\begin{eqnarray}\label{29}
S_{\Phi}&\simeq&\frac{A}{4G}-\frac{9}{2}\alpha
L_p^2\left(\frac{4\Omega_4}{4G}
\right)^{\frac{2}{3}}({\frac{A}{4G}})^{\frac{1}{3}}
-3\left(-\frac{5}{32}\alpha'+\frac{23}{128}\alpha^2\right)L_p^4
\left(\frac{4\Omega_4}{4G}\right)^{\frac{4}{3}}\left(\frac{4G}{A}
\right)^{\frac{1}{3}}
\nonumber\\&-&\left(-\frac{7}{128}\alpha''+\frac{37}{256}\alpha
\alpha'-\frac{273}{3072}\alpha^3\right)L_p^6\left(\frac{4\Omega_4}{4G}\right)^2\left(\frac{4G}{A}\right)\,,
\end{eqnarray}

\begin{eqnarray}\label{30}
S_{\Phi}&\simeq&\frac{A}{4G}-\frac{6}{8}\alpha
L_p^2\left(\frac{5\Omega_5}{4G}
\right)^{\frac{1}{2}}({\frac{A}{4G}})^{\frac{1}{2}}
+\left(-\frac{5}{32}\alpha'+\frac{23}{128}\alpha^2\right)L_p^4
\left(\frac{5\Omega_5}{4G}\right)\ln{\frac{A}{4G}}
\nonumber\\&-&2\left(-\frac{7}{128}\alpha''+\frac{37}{256}\alpha
 \alpha'-\frac{273}{3072}\alpha^3\right)L_p^6\left(\frac{5\Omega_5}{4G}
 \right)^{\frac{3}{2}}\left(\frac{4G}{A}\right)^{\frac{1}{2}}\,.
\end{eqnarray}
respectively. The entropy for a FRW universe with other
dimensionality can be similarly derived. It is seen that the
logarithmic correction term appears only for odd $n$ (even spacetime
dimensionality). Now, one can compare the entropy of the FRW
spacetime calculated in the GUP and MDR frameworks by asking if they
are functionally equivalent.

Comparison with equations (\ref{15}-\ref{17}) shows that the results
of extra dimensional form of the GUP and MDR as functions of the
apparent horizon and planck length are functionally the same, apart
from numerical coefficients. Since GUP and MDR have essentially the
same phenomenology, having the results with similar functional form
is reasonable. Therefore, one may conclude that our suggested form
of the MDR for a model universe with extra dimensions is in fact
reasonably satisfactory. It is also necessary to point out that the
Planck length in the extra dimensional form of the MDR depends on
the dimensionality of the spacetime. In addition, the Planck length
appearing in the expressions for GUP and MDR in 4-dimensions are the
same \cite{30}. The Functional consistency between the results of
extra dimensional form of GUP and MDR for the variables $A$ and
Planck length and the necessity of having the same result in each
approach points to the possibility that the Planck length in the
extra dimensional forms of the GUP and MDR is also the same, that is
$l_{p}$ of section 2.1 is equivalent to $L_{p}$ in  2.2.

As another point, we can mention the logarithmic correction term in
the 4-dimensional spacetime. The emergence of a positive logarithmic
correction term within the MDR and GUP in 4-dimensional FRW
spacetime is interesting to note. As in our earlier work \cite{30},
the parameter $\alpha$ in MDR is a negative quantity of order one
(see also \cite{31}). There, we compared the results of two
approaches, the generalized uncertainty principle and modified
dispersion relation within the context of black hole thermodynamics
with that of the string theory and Loop quantum gravity. Demanding
the same results in all approaches and considering string theory and
loop quantum gravity as more comprehensive, we put some constraints
on the form of GUP and MDR. Also, we found that GUP and MDR are not
independent concepts. In fact, they could be equivalent in an
ultimate quantum gravity theory. The existence of a positive minimal
observable length necessitates a positive value for the model
dependent parameter $\alpha$ in the form of GUP. Since we know the
relation between the model dependent parameters in GUP and MDR in
\cite{30}, we set the parameter $\alpha$ as a negative value for MDR
in this paper. Now, it is easy to see that the logarithmic
correction term in a 4-dimensional FRW spacetime is positive in both
approaches. It is interesting to note that the existence of a
logarithmic term in the entropy-area relation is restricted to the
even dimensionality of the spacetime  within both GUP and MDR. This
result may be helpful in paving the way for a better understanding
of quantum gravity.

\section{Conclusions}
In this work, we obtained the entropy of a FRW universe for
different dimensions of spacetime via the well-known extra
dimensional form of the generalized uncertainty principle. We also
suggested a form for the modified dispersion relation in a model
universe with extra dimensions in the same way as that in
$4$-dimensions, except that the Planck length would now depend on
the dimensionality of the spacetime. This is actually the case since
one expects that the fundamental length scale in MDR, $L_p$, would
be different from that in 4-dimensions in order to account for the
the existence of the extra dimensions. By this assumption, we
obtained the entropy of a FRW spacetime within the extra dimensional
modified dispersion relation formalism. Since GUP and MDR have
essentially the same phenomenology, they must lead to results
exhibiting the same functional form, but not necessarily the same
numerical coefficients. We found that the results have equivalent
functional forms in terms of $\frac{A}{4G}$. In other words, the
entropy of the FRW universe calculated based on the GUP and our
suggested MDR are essentially the same. Since the form of the GUP
for a model universe with extra dimensions is known, this
equivalency shows that our suggested form of the MDR is conceivable.
The GUP and MDR being equivalent may therefore be looked upon as the
common feature of all quantum gravity theories. Having a
functionally equivalent entropy in these two models, leads to having
a relationship between model dependent parameters in GUP and MDR
which may be considered as another interesting outcome of our study.
One may also assume that the Planck length expression in extra
dimensional form of the GUP and MDR is the same. On the other hand,
we have found that the logarithmic correction term, whose existence
is still somewhat mysterious, may only emerge in the
even-dimensional FRW universe in both GUP and MDR models. Therefore,
the existence of the logarithmic correction term in the entropy
formulae depends on the dimensions of the spacetime. If one insists
on the existence of the logarithmic correction term in the entropy
of a FRW universe, the spacetime dimensions is restricted to be
even. This fact provides a constraint on any viable quantum gravity
theory.

Another point that should be stressed here is the connection between
MDR and the spacetime non-commutativity. In fact, MDR is a
manifestation of the spacetime non-commutativity in quantum gravity.
Therefore, our knowledge of the  MDR in spacetimes with extra
dimensions provides a background to study non-commutativity in extra
dimensions. It seems that a complete knowledge of the form of the
MDR in model universes with extra dimensions opens new directions in
quantum gravity and noncommutative geometry.

\end{document}